\documentclass[preprintnumbers,amsmath,amssymb]{revtex4}

\usepackage{dcolumn}
\usepackage{bm}

\usepackage{graphicx}

\newcommand{\msq}{m_{\tilde{q}}}
\newcommand{\mgo}{m_{\tilde{g}}}
\newcommand{\msg}{m_G}
\newcommand{\sq}{\tilde{q}}
\newcommand{\go}{\tilde{g}}

\newcommand{\ptj}{p_{T,j}}

\def\ie{\hbox{\it i.e.\,}{}}  

\def\eg{\hbox{\it e.g.\,}{}}

\def\Tr{\mathop{\rm Tr}}

\relax

\begin{document}

\preprint{ANL-HEP-PR-08-65,
          Edinburgh 2008/43,
          NU-HEP-TH/08-08}

\title{Seeking Sgluons}

\author{Tilman Plehn}
\affiliation{SUPA, School of Physics and Astronomy, 
             University of Edinburgh, Scotland}

\author{Tim M.P. Tait}
\affiliation{Argonne National Laboratory, Argonne, IL 60439 \\ 
             Northwestern University, 2145 Sheridan Road, Evanston, IL 60208}

\begin{abstract}
Scalar gluons --- or sgluons --- are color octet scalars without
electroweak charges. They occur in supersymmetric models of Dirac
gauginos as the scalar partners of the gluino and carry
Standard-Model type $R$ charge. This allows them to interact with
ordinary matter and to be produced at the LHC, singly as well as in
pairs. Sgluons dominantly decay into gluons, top pairs, and a top
quark plus a light quark. A pair of sgluons decaying into like-sign tops
would provide a striking signature at the LHC. In our discussion of this
channel we especially focus on the proper treatment of QCD jets.
\end{abstract}

\maketitle

\section{Introduction}
\label{sec:intro}

Theories with weak scale supersymmetry represent the most complete and
cherished vision of physics beyond the Standard Model.  Their
many successes include stabilization of the electroweak scale with
respect to high-scale physics, improvement of the convergence of
couplings necessary for Grand Unification, possible electroweak
baryogenesis to explain the matter-anti-matter asymmetry of the
Universe, and (with $R$ parity) relatively mild contributions to
precision electroweak data and a successful dark matter
candidate.\medskip

However, none of those successes rest crucially on the minimal
realization of the supersymmetric Standard Model~\cite{Kribs:2007ac}.  In fact, there are
features of the minimal supersymmetric standard model (MSSM) which are
somewhat at odds with the MSSM as a completely natural theory of the
electroweak scale.  The $\mu$ or $B\mu$ problem indicates that the
supersymmetric $\mu$ term must be roughly of the same size as the
supersymmetry-breaking parameters, and yet the MSSM offers no
explanation for why this should be the case. The lightest Higgs mass
in the MSSM is, at tree-level, less than the $Z$ boson mass, and even
including radiative corrections typically in conflict with the LEP-II
bound~\cite{RuhlmannKleider:2007zz}. The model survives based on large radiative
corrections from scalar top quarks, but in turn this requires the stops to be so
heavy that the Higgs soft masses end up fine-tuned to the per-cent
level.

Perhaps the most disturbing feature of the MSSM is the fact that if
one naively assumes a general spectrum of supersymmetry-breaking
parameters, it is simply ruled out.  If there are large mixings
among the squarks and among the sleptons, flavor-violating processes such as
$K$-$\overline{K}$ mixing, $\mu \rightarrow e \gamma$, and others, can
be enhanced by orders of magnitude with respect to Standard Model
predictions, in obvious contradiction to experimental data.  The
traditional solution (of which gauge mediation is the prototype) is to
engineer supersymmetry breaking such that the soft-breaking mass
parameters are highly flavor diagonal~\cite{susy_flavor}.  This
controls flavor violation to an acceptable level at a given energy
scale, but continues to be challenged by natural electroweak symmetry
breaking, the Higgs mass, and cosmology.\medskip

An alternative is to alter the low energy structure of the model such that the
supersymmetric solution to the hierarchy problem is preserved, but flavor
violation is ameliorated.  A recent elegant solution imposes a continuous $R$
symmetry~\cite{Hall:1990hq}, to form the minimal $R$-symmetric supersymmetric Standard Model
(MRSSM)~\cite{Kribs:2007ac}.  This model invokes the $R$ symmetry to forbid
the potentially flavor-violating left-right mixing $A$-terms, and to guarantee
(at least largely in light of anomaly-mediated contributions) Dirac gaugino
masses. This Dirac nature often leads to an additional suppression in
flavor-violating processes from gauginos running in the loops, which causes
many contributions to scale with large gaugino mass as $1 / \mgo^2$ instead of
the Majorana scaling, $1/\mgo$ ~\cite{Kribs:2007ac,Blechman:2008gu}.  The
model has additional interesting features, such as UV finite scalar masses due
to the super-soft feature of Dirac gauginos~\cite{Fox:2002bu}, and additional
matter content at the electroweak scale. For example, if the gluino is a Dirac
particle, the two on-shell degrees of freedom from the gluon are not
sufficient to construct a supersymmetric four-spinor. To provide the two
additional bosonic degrees of freedom, the MRSSM adds a complex scalar
partner. An unbroken $R$ symmetry is also a generic feature of simple models
of meta-stable supersymmetry breaking~\cite{Amigo:2008rc}.\medskip

In this article we consider the phenomenology of the scalar partner of the
Dirac gluino, a color adjoint scalar --- the sgluon.  As a colored particle,
it couples to a gluon and will have large pair-production rates at the
Tevatron and LHC. It interacts with quarks at the one-loop level proportional
to the quark masses. Thanks to the large mixing naturally expected in the
MRSSM's squark sector, it will readily decay into flavor violating channels.
Previous studies have considered scalar color octets with either decays into
missing energy~\cite{Dobrescu:2007xf}, with electroweak as well as $SU(3)$
charges~\cite{Gerbush:2007fe}, or purely flavor diagonal
couplings~\cite{Dobrescu:2007yp}, which all lead to very different
phenomena.\medskip

Our work is organized as follows.
In Section~\ref{sec:sgluons} we show how sgluons arise in a model with
Dirac gluinos, examine their soft masses, and derive their
interactions including the one-loop contribution to the
$G$--$q$--$\bar{q}'$ and $G$--$g$--$g$
vertices. In Section~\ref{sec:fcnc} we 
examine the flavor-violating effects mediated by sgluons and
derive some mild constraints from $K$-$\bar{K}$ mixing.
In Section~\ref{sec:search} we discuss the production of a
pair of sgluons through the strong interaction, and examine the decay
of the pair into like-sign tops. This distinctive signature provides
bounds on the sgluon mass at the Tevatron and will easily be
discovered at the LHC for a wide range of masses. We particularly
focus on the proper treatment of the jet activity in such events.

\section{Sgluons}
\label{sec:sgluons}

Scalar gluons are contained in a chiral superfield $\Phi^a$ which is a
color adjoint carrying $R$ charge zero.  The fermionic component
$\psi^a$ is married through $D$--term
supersymmetry breaking to the ordinary gluino $\lambda^a$. The
lowest component $G^a$ is a complex color adjoint scalar.
Supersymmetry breaking will generally split this into two real scalar
states which are admixtures of the real and imaginary parts of $G$.
We discuss the spectrum below.\medskip
 
Kinetic terms for the sgluons are contained in canonical K\"{a}hler
potential terms for $\Phi$,
\begin{equation}
 \int d^4 \theta  ~~ \Phi^\dagger e^{-V} \Phi
\end{equation}
where $V$ is the vector superfield containing the gluon and the $SU(3)_C$
gauge indices are implied.  The kinetic terms include the coupling of the
sgluon $G$ to the gluons from the covariant derivative, a $G$-$\psi$-$\lambda$
coupling of strength $g_s$ required by supersymmetry, and $D$--term
contributions to the scalar potential that are of the form
$G^*$-$G$-$\sq^*$-$\sq$ which will not be important for our purposes.  There
are no renormalizable gauge invariant terms through which $\Phi$ interacts
with matter superfields in either the K\"{a}hler potential or in the
super-potential, and the assumed $R$ symmetry is incompatible with $\Phi^2$ or
$\Phi^3$ interactions in the super-potential.  Thus, the tree-level
supersymmetric interactions of $G$ are determined entirely by supersymmetric
QCD,
\begin{equation}
 {\cal L}_{\rm SQCD} = \left( D_\mu G \right)^* 
            \left( D^\mu G \right) 
          + i \sqrt{2} \; g_s f_{abc} \; 
            \bar{\go}^b \left( G^a P_L + G^{a*} P_R \right) \go^c 
\label{eq:susy_qcd}
\end{equation} 
where $D^\mu$ is the usual covariant derivative for a color adjoint,
$\go$ is the (four-component) gluino, and $f_{abc}$ are the
structure constants of $SU(3)$ .

\subsection{Supersymmetry breaking and masses}

Soft mass terms for the sgluons can arise from either $F$--term
($\langle X \rangle = \theta^2 F$) or $D$--term ($\langle W^\prime
\rangle = \theta D^\prime$) spurions of supersymmetry breaking,
\begin{equation}
  \int d^4 \theta \left\{ \frac{1}{M_1^2} X^\dagger X \Phi^\dagger \Phi 
                        + \frac{1}{M_2^2} X^\dagger X \Tr \Phi^2
                  \right\}
+ \int d^2 \theta \frac{1}{M_3^2} \;
           W^\prime_\alpha W^{\prime \alpha} \Tr \Phi^2 + \text{H.c.}
\end{equation}
and also are generated by the term responsible for the Dirac gluino
mass~\cite{Fox:2002bu},
\begin{equation}
  \int d^2 \theta \; \frac{\sqrt{2}}{M_4} 
                 \; W^{\prime \alpha} W^{a}_{3 \alpha} \; \Phi^a
\end{equation}
where $W^{a}_{3 \alpha}$ is the usual superfield $SU(3)_C$ field strength.  
This super-potential term, along with the usual MSSM $SU(3)_C$ $D$-terms,
lead to terms in the Lagrangian,
\begin{equation}
-\mgo \lambda^a \psi^a - \sqrt{2} \left( \mgo G^a + \mgo^* G^{a *} \right) D^a
- g_s D^a  \sum_{\sq_L}  \sq_L^{*} T^a \sq_L 
+ g_s D^a  \sum_{\sq_R}  \sq_R T^a \sq_R^{*}
-\frac{1}{2} D^a D^a
\end{equation}
where $\mgo = D^\prime / M_4$ is the Dirac gluino mass, 
and $T^a$ are the generators
of $SU(3)_C$ in the fundamental representation. 
Replacing the $SU(3)$ auxiliary field $D^a$ through its 
equation of motion leads to terms proportional to
$\mgo^2 G^2$ and $\mgo^{*2} G^{* 2}$
as well as $|\mgo|^2 |G^2|$~\cite{Kribs:2007ac}.  It also induces
tri-linear interactions of $G$ with squarks, somewhat analogous to
the $A$ terms in the usual MSSM.  Altogether, the
supersymmetry-breaking Lagrangian for $G$ reads
\begin{equation}
{\cal L}_{\rm soft} = m_1^2~ | G^a |^2 
                  + \frac{1}{2} m_2^2~ G^{a 2} 
                  + \frac{1}{2} m_2^{* 2}~ G^{a * 2}
                  - \sqrt{2} g_s \left( \mgo G^a 
                                       +\mgo^* G^{a *} \right) 
                    \; 
                    \left( \sum_{\sq_L}  \sq^{*}_L T^a \sq_L - \sum_{\sq_R} \sq_R T^a \sq_R^{*}  \right)
\label{eq:soft}
\end{equation}
where $m^2_1$ is a real parameter and $m^2_2$ may be complex.
The mass eigenstates are two real color adjoint scalars which can be
labelled as $G_1$ and $G_2$.  The mass-squared
eigenvalues are given by,
\begin{equation}
 m^2_{G_1,G_2} = m^2_1 \mp | m^2_2 | ~,
\end{equation}
and clearly we must have $m^2_1 > | m^2_2 |$ or run the risk of a
color-breaking vacuum.  There will be a non-trivial mixing angle when
$m^2_2$ is complex.  When we write $m^2_2$ in terms of its phase
$m^2_2 = |m^2_2| e^{i \gamma}$, the mass eigenstates are
\begin{alignat}{5}
G^a_1   = & \sin \frac{\gamma}{2}~G^a 
             + \cos \frac{\gamma}{2}~G^{a *} ~, \notag \\
G^a_2   = & \cos \frac{\gamma}{2}~G^a 
             - \sin \frac{\gamma}{2}~G^{a *} ~.
\end{alignat}
For simplicity we will assume $\mgo$ and $m^2_2$ are
real, and thus there is no non-trivial mixing from here on.
In that case both sgluons are either a pure scalar or a pure pseudoscalar, 
which is equivalent as long as we combine them with massless QCD, the
theory relevant for LHC.

All tree-level interactions of the sgluon with Standard Model and MSSM
states we can read off $\mathcal{L}_{\rm SQCD}$ and
$\mathcal{L}_{\rm soft}$. The coupling of two sgluon to gluons is
simply a result of its adjoint color charge and arises from the
kinetic term. The coupling to two gluinos is the supersymmetric
partner of the gluon couplings, while the couplings to two squarks
arise from $D$ terms. Note in particular  that the Dirac gluino mass
sets the size of the squark-squark-sgluon coupling.

\begin{figure}[t]
\begin{center}
  \includegraphics[angle=0,scale=0.25]{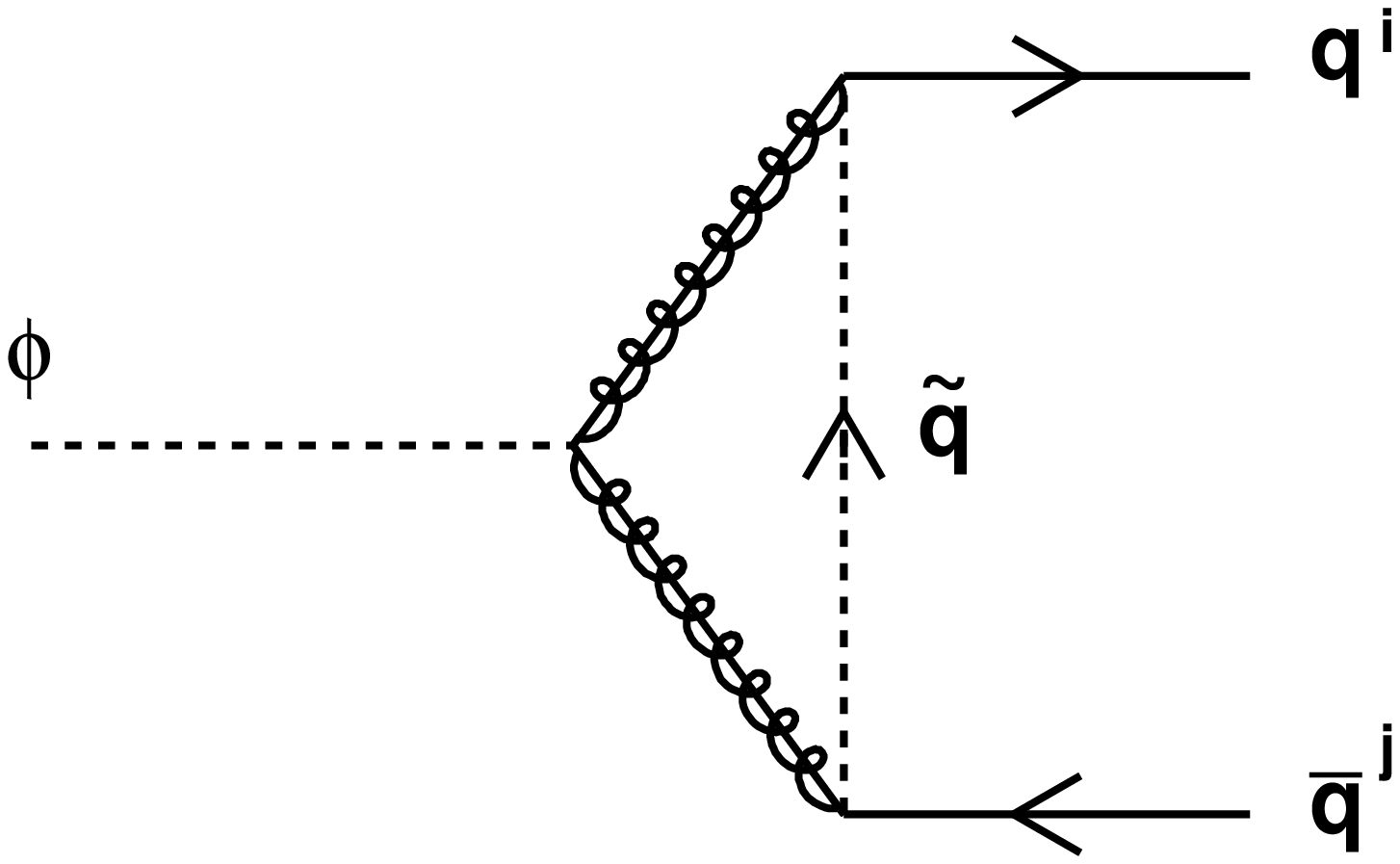} \hspace*{1.5cm}
  \includegraphics[angle=0,scale=0.25]{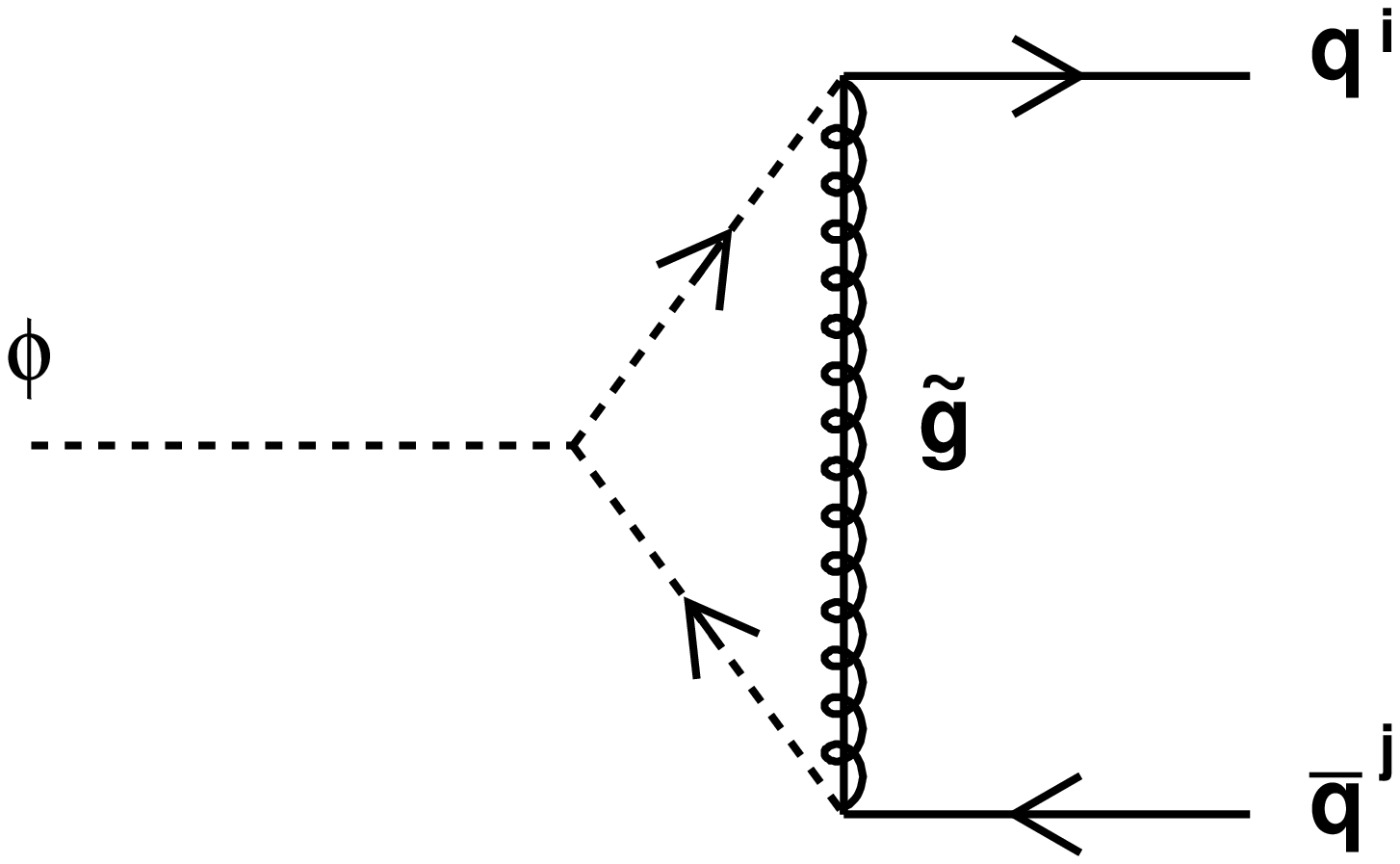} 
\end{center}
\caption{\label{fig:qvertex} Feynman diagrams for sgluon interactions
  with quarks.}
\end{figure} 

\subsection{Loop-induced coupling to quarks}
\label{sec:qvertex}

If we insert the squark and gluino couplings to a sgluon shown in 
eq.(\ref{eq:susy_qcd}) and in eq.(\ref{eq:soft}) into one-loop diagrams,
the sgluon will couple to quarks. There are
two Feynman graphs responsible for this interaction induced by gluinos
and squarks, shown in Figure~\ref{fig:qvertex}. 
For each of the two sgluon states the effective action
after electroweak symmetry breaking contains a dimension-4 operator of the
form,
\begin{equation} 
 G^a \; \left[ \, \bar{q}^j \; T^a \, \left( g^{ij}_L P_L 
                                              + g^{ij}_R P_R \right)  
                    q^i \, \right] + \text{H.c.}
\end{equation}
where $i$ and $j$ are quark flavor indices.
The couplings may be expressed as
\begin{equation}
g^{ij}_{(L/R)}  =  
 \frac{g_s^3 \widetilde{\delta}^{(L/R)}_{ij}}{16 \pi^2} \frac{\mgo}{\msg^2}
  \left( m_i f^{(L/R)}_{q,i} 
       - m_j f^{(L/R)}_{q,j} \right)
\label{eq:qvertex}
\end{equation}
where $\widetilde{\delta}_{ij}$ is the relevant squark mixing
parameter, $m_{i,j}$ are the quark masses, and the dimensionless $f_q$s
are functions of the heavy sgluon, gluino, and the squark
masses. Their form is given in the Appendix.  It is important to
notice that in the MRSSM these couplings come out
automatically proportional to
quark masses, which mitigates their contribution to
flavor-violating observables. In contrast, for
example, the $A$ terms in the MSSM have to be {\em defined} to appear
proportionally with the quark masses, \ie as $m_q A_q$, which is an
additional assumption on the flavor structure of the general MSSM.  In
the limit of degenerate squarks, this source of flavor violation in the MRSSM
will switch off through a super-GIM mechanism. In the limit of large squark,
gluino, or sgluon masses the quark-quark-sgluon coupling will be suppressed by
$\mgo / \msq^2$, $1 / \mgo$, or $\mgo / \msg^2$, respectively.

\begin{figure}[t]
\begin{center}
  \includegraphics[angle=0,scale=0.25]{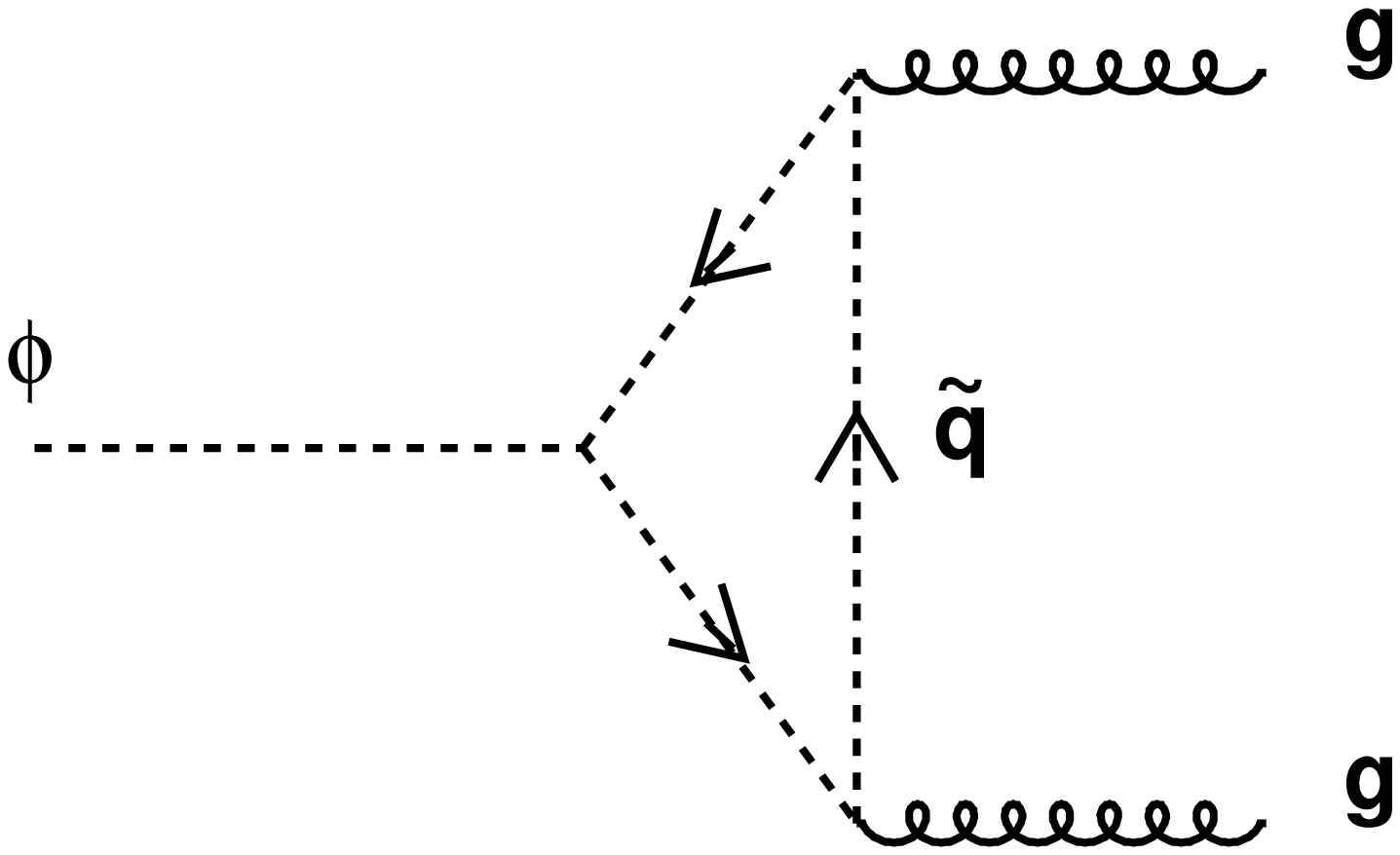} \hspace*{0cm}
  \includegraphics[angle=0,scale=0.25]{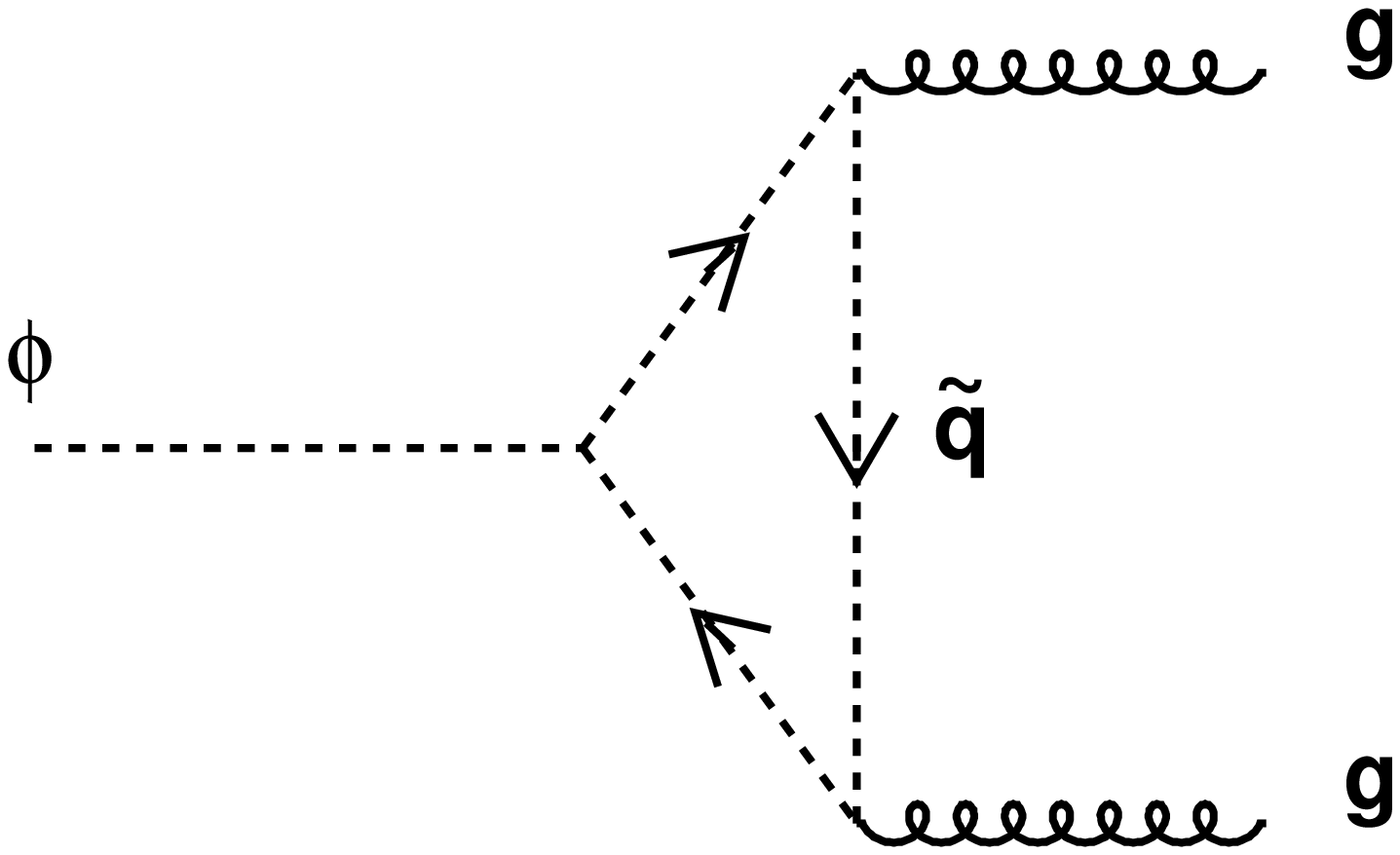} \hspace*{0cm}
  \includegraphics[angle=0,scale=0.25]{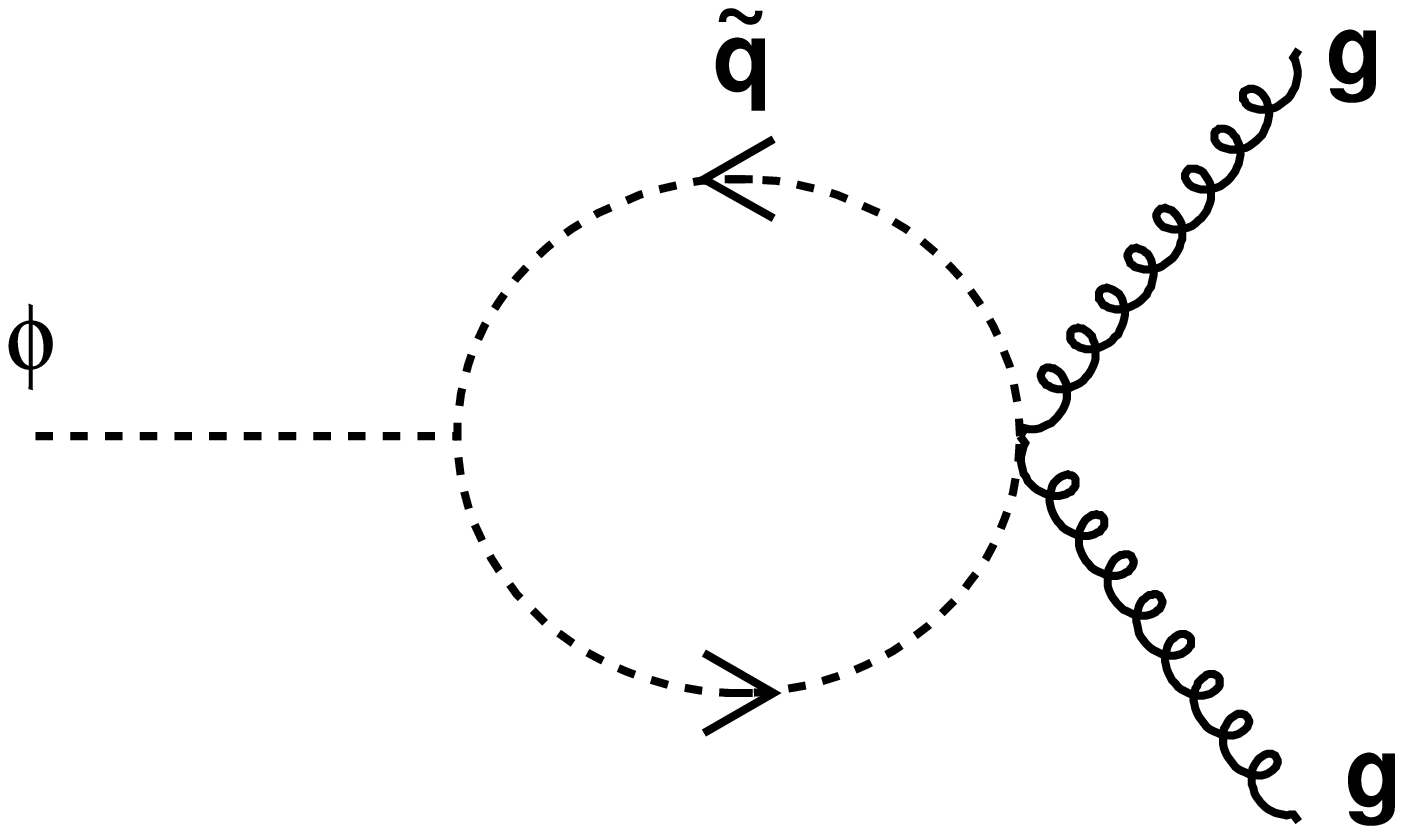}
\end{center}
\caption{\label{fig:gvertex} Feynman diagrams for single sgluon
  interactions with gluons.}
\end{figure} 

\subsection{Sgluon-gluon-gluon Coupling}
 
Pairs of sgluons interact with one or two gluons as a consequence of the fact
that the sgluons are adjoints of $SU(3)$.  There is also a single
sgluon-gluon-gluon interaction mediated by squarks, as shown in
Figure~\ref{fig:gvertex}.  The effective action contains a dimension-five
operator generated at one loop
\begin{equation}
 \frac{g_s^3}{16 \pi^2} \; \frac{\mgo}{\msg^2} \; d^{abc} \; 
 \lambda_g \; G^a F^b_{\mu \nu} F^{c \mu \nu} + \text{H.c.}
 \label{eq:sggg}
\end{equation}
where $d^{abc}$ is the symmetric gauge-invariant combination of three
$SU(3)$ adjoints and the dimensionless form factor $\lambda_g$ is
given in the Appendix.  A similar contribution mediated by gluinos
vanishes at one loop, due to the symmetry structure of the (asymmetric)
color factor and the (symmetric) loop contribution. In the limit of
heavy squarks the coupling $\lambda_g$ will vanish proportionally to
$1/\msq^2$.

\section{Sgluon-mediated flavor violation}
\label{sec:fcnc}

Through its flavor-violating couplings to quarks, the sgluon can also
mediate flavor-changing processes.  At energies far below the sgluon
mass, these interactions look like $\Delta F=2$ four fermion
interactions.  Using the $s$-$d$ transition operators relevant for
$K$-$\overline{K}$ mixing we can illustrate the terms left behind when
the sgluon is integrated out:
\begin{equation}
-\frac{1}{2 \msg^2} \left\{ \left( g^{sd~2}_L + g^{sd~2}_R \right) 
                            \left[ Q^{sd}_5 - \frac{1}{3} Q^{sd}_4 \right]
                  + g^{sd}_R g^{sd}_L \left[   Q^{sd}_3 
                                           + \widetilde{Q}^{sd}_3
                                           -\frac{1}{3} Q^{sd}_2 
                                           - \frac{1}{3} \widetilde{Q}^{sd}_2  
                                    \right]
                    \right\}
\end{equation}
where the $Q^{ij}_i$ are the four fermion interactions defined in
Ref.~\cite{Bona:2007vi}. Comparing to eq.(\ref{eq:qvertex}) we know that the
couplings $g$ are proportional to $\widetilde{\delta}$, so the relevant
parameters which will be constrained are $\widetilde{\delta}/\msg$.  Analogous
expressions describe $b$-$d$, $b$-$s$, and $c$-$u$ mixing.  We compute the
coefficients of each of these operators and compare with the global fit by the
UT$fit$ collaboration~\cite{Bona:2007vi} to obtain bounds on the sgluon mass,
for a given choice of gluino mass, average squark masses, and squark flavor
mixing parameters $\widetilde{\delta}$.\medskip

\begin{figure}[t]
\begin{center}
  \includegraphics[width=7cm]{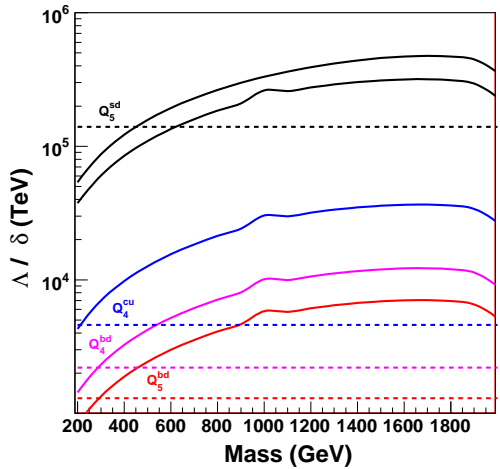} 
\end{center}
\caption{Solid lines: effective scales characterizing the four fermion
  operators mediated by sgluons
  contributing to (bottom to top) $B_d$ mixing (2 curves), $D$ mixing,
  $K$ mixing with $\mgo = 1$~TeV and an average $\msq = 500$~GeV, and
  $K$ mixing with $\mgo = \msq = 1$~TeV.  Dashed lines: current bounds
  on the operators (bottom to top), $Q_5^{bd}$, $Q_4^{bd}$,
  $Q^{cu}_4$, and $Q^{sd}_5$ (imaginary part),
  obtained from the global fit  \protect{\cite{Bona:2007vi}}.
\label{fig:fcnc}}
\end{figure}

In Fig.~\ref{fig:fcnc}, we present the scale of the most constraining of the
four fermion operators, expressed as the effective scale $\Lambda /
\widetilde{\delta}^{ij}$.  Since mixing effects will only raise the effective
scale of the four fermion interactions, the curves are the minimum possible
effective scales for a given choice of sgluon, gluino, and average squark
masses.  In other words, a realistic choice of mixing parameters will increase
the effective scales which suppress the FCNC operators and thus result in less
constraints from flavor violation mediated by sgluons.  We also show the
bounds on the corresponding scales from Ref.~\cite{Bona:2007vi}. Bounds from
$B$-$\bar{B}$ and $D$-$\bar{D}$ mixing are mild enough as to basically provide
no constraint. The $K$-$\bar{K}$ mixing bound is more severe, but the plotted
bound is on the imaginary part of $\tilde{\delta}$, and thus can be avoided if
there is no large $CP$-violating phase in the down-strange squark mixing
element.  In any case, sgluon masses above 600 GeV or so are compatible with
measurement, regardless of mixing.  Note also that the constraints on
down-type mixing do not in any way preclude the sgluon flavor-violating decays
into up-type quarks which we consider below.

\section{Sgluons at colliders}
\label{sec:search}

As discussed in Section~\ref{sec:sgluons}, we expect the
$R$--symmetric supersymmetric theory to include a pair of sgluons
whose mass is split by something of order the gluino mass. There will
thus be two sgluon states which we can search for at hadron colliders.
Since the masses are not typically degenerate, the results in this
section are presented for a single sgluon state, and apply equally to
the lighter or the heavier sgluon.
The couplings to quarks and single sgluon coupling to gluons will
depend on the mixing between the two states, but tree level pair
production of sgluons involves only the strong coupling, as these
interactions are protected by $SU(3)_C$ gauge invariance.  The
resulting pair cross sections thus only depend on the sgluon mass, similar
to, for example, the case of scalar leptoquark pairs.\medskip

\begin{figure}[t]
\begin{center}
  \includegraphics[width=8cm]{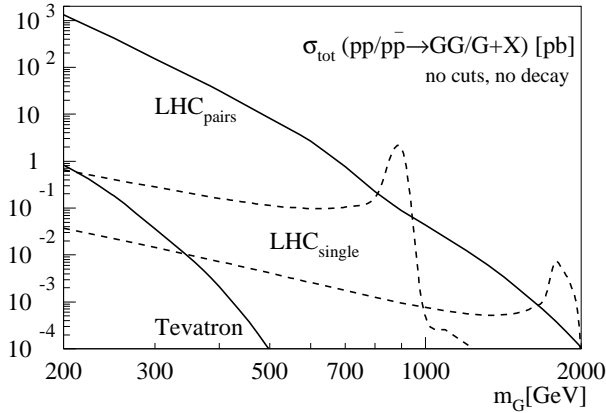} 
\end{center}
\caption{Inclusive production cross sections for sgluons at the
  Tevatron and at the LHC. For the LHC we show pair production (solid)
  and single production (dashed) as a function of the sgluon mass. The
  two curves for single sgluon production assume a gluino mass of
  1~TeV and squark masses of 500~GeV (upper curve) and 1~TeV
  (lower curve).
\label{fig:xsec}}
\end{figure} 

In Figure~\ref{fig:xsec}, we present the leading order cross section
for pair production of sgluons as a function of their mass at the
Tevatron and LHC~\cite{Alwall:2007st}. For masses around 250~GeV, the
production rate of the order of 200~fb at the Tevatron would correspond
to a few hundred sgluon pair events in the currently available CDF and
DZero data, depending on the decays and triggers.  As expected, the
production rate at the Tevatron drops below the femtobarn level for
sgluon masses around 400~GeV, thanks to the limited center-of-mass
energy and $p$--wave suppression of the dominant subprocess $q \bar{q}
\rightarrow G G$. From a dedicated analysis we could expect
sgluon mass bounds similar to squark mass bounds in the limit of
large gluino mass.\medskip

At the LHC, the dominant subprocess is $gg \rightarrow G G^*$ with large cross
sections, falling from around 400~pb for masses around 250~GeV to 2~fb for
masses around 1.5~TeV. For sgluon masses in the TeV range, the LHC will rely
on its sea-quark suppressed $q \bar{q}$ luminosity which leads to a rapid drop
of the cross section above 1.8~TeV.  With an appreciable LHC luminosity these
rates correspond to several hundred to a few million events available for
analyses.

For the LHC we also present the single production rate from gluon
fusion~\cite{Spira:1995mt}, for two choices of squark and gluino masses.
Through the one-loop diagrams discussed in Sec.~\ref{sec:sgluons} there can be
appreciable single sgluon production through $gg \rightarrow G$, which can
dominate for large sgluon masses because of the phase space suppression of the
pair production and threshold effects. For small sgluon masses the LHC is not energy limited, so the
single production channel is suppressed by a loop factor $\alpha_s/(4 \pi)$
squared. The problem of single production will be challenging backgrounds
discussed below.  The sgluon interaction with quarks typically results in a
negligible single sgluon production rate, because the leading contribution is
one loop and the suppression by the light quark masses.\medskip

Sgluons can also be produced at the LHC in cascade decays involving squarks,
through the soft breaking interaction of eq.(\ref{eq:soft}).  Production of
pairs of heavier squarks can thus cascade down through sgluons into lighter
squarks, leaving behind either light jets which reconstruct the sgluon mass,
or events enriched with top quarks.  Either of these possibilities is rather
exotic from the point of view of the standard MSSM, and we leave their
detailed exploration for future work.

\begin{figure}[t]
\begin{center}
  \includegraphics[angle=0,scale=0.33]{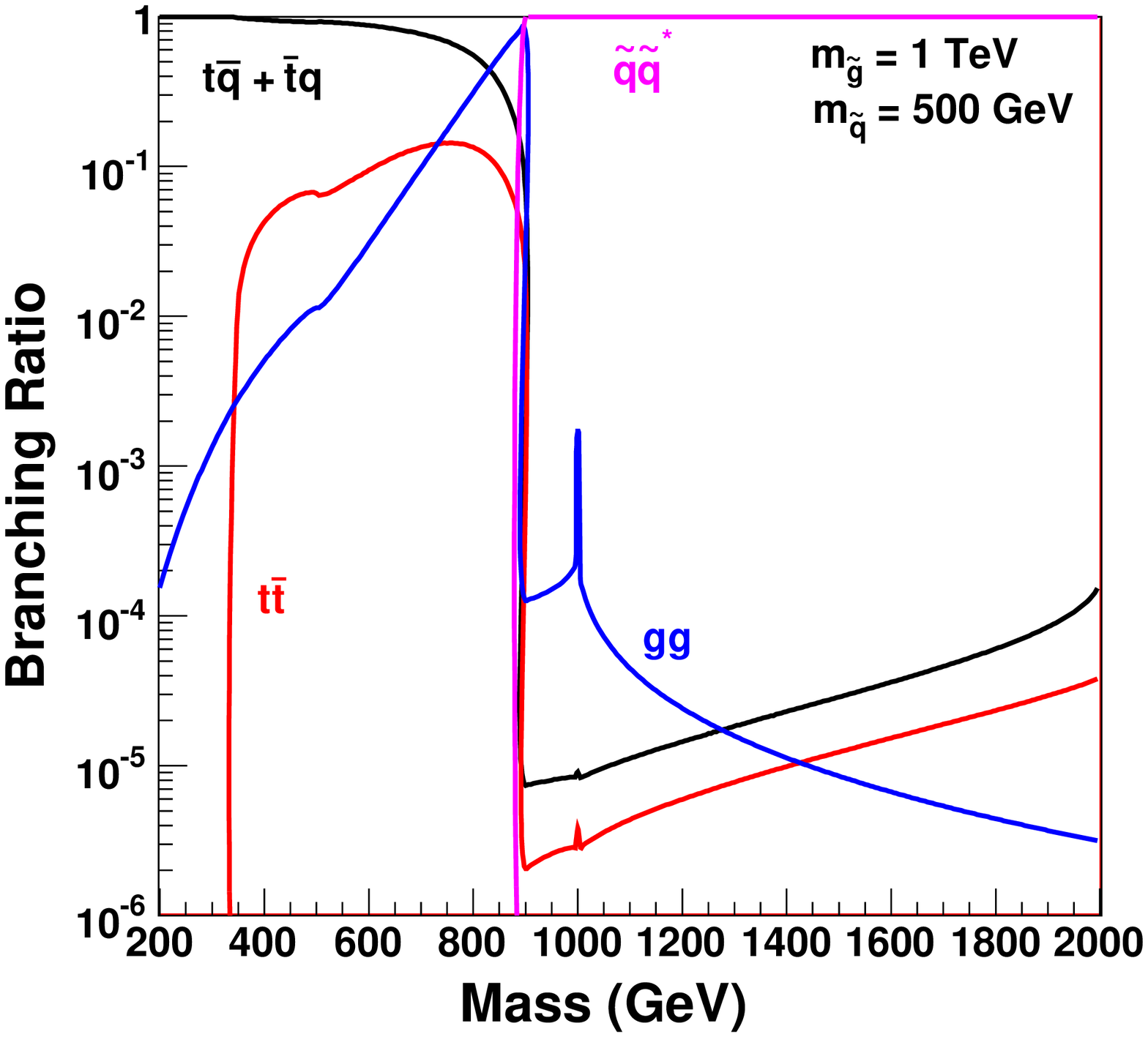} \hspace*{1.5cm} 
  \includegraphics[angle=0,scale=0.33]{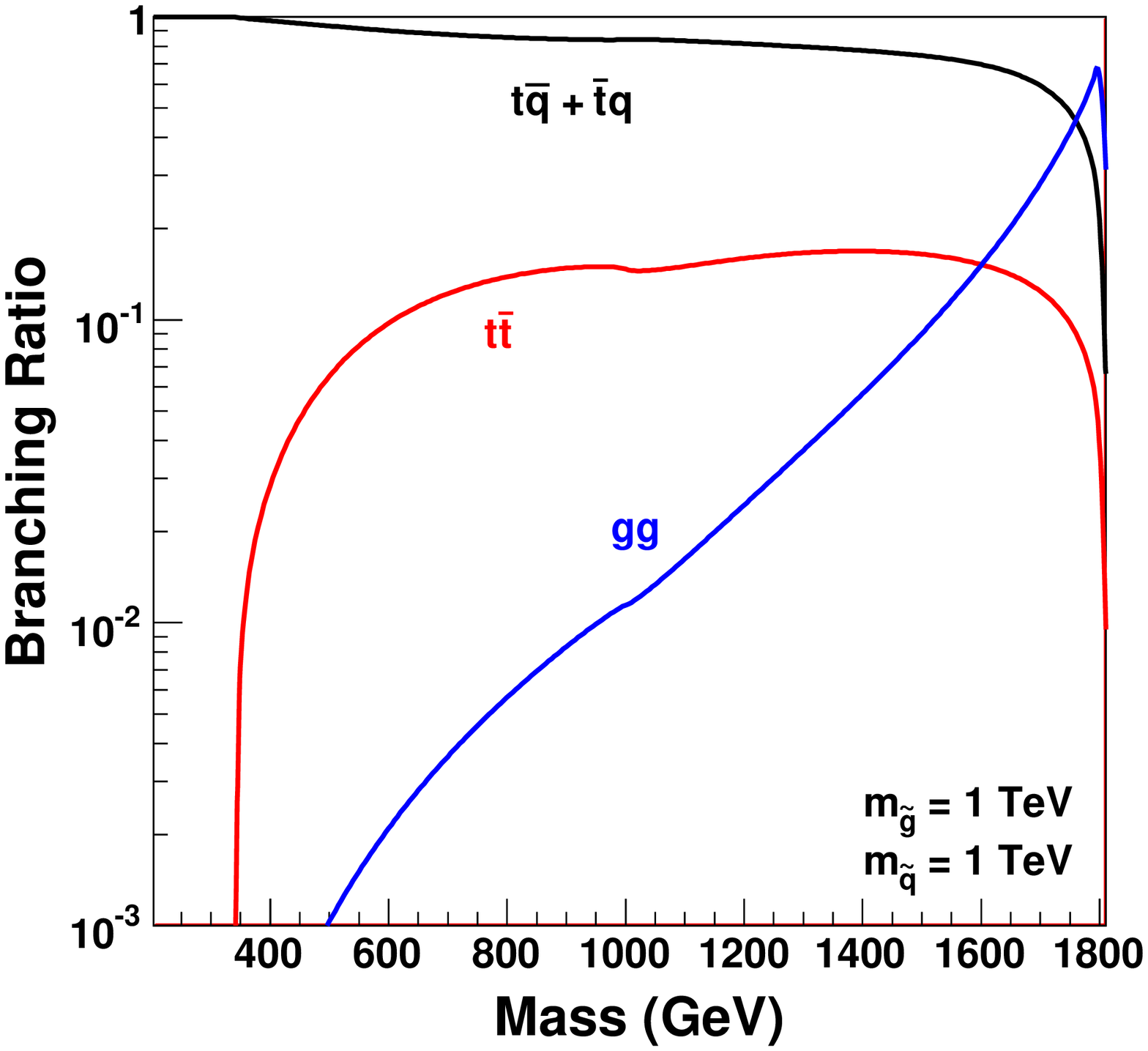}
\end{center}
\caption{\label{fig:br}Branching ratios for sgluon decays into $gg$,
  $t\bar{t}$, $\bar{q}t+\bar{t}q$ for $q=u,c$ and $\sq
  \sq^*$ as a function of sgluon mass and for two choices of
  left-handed squark and gluino masses.  Right-handed squark masses 
  are set to $90\%$ of the left-handed squark masses.
  We assume maximal up-squark
  mixing.}
\end{figure} 

\subsection{Sgluon decays --- like-sign tops}
\label{sec:decays}

If heavy enough, sgluons will decay at tree level into pairs of
gluinos and/or squarks.  If these decay channels are closed, the
sgluon has to decay through its loop-induced couplings into quarks and
gluons. In the case of decays into quarks, the $G$-$q$-$\bar{q}$
interaction is proportional to the heavier of the two quark masses, so one
can expect that decays including at least one top quark to dominate.

If the mixing in the up-type squark sector is large (as is the point of the
MRSSM), we therefore expect large and comparable branching ratios into $t
\bar{u}$, $t \bar{c}$, and $t \bar{t}$. The specifics of the branching ratios
are a window into the details of the squark mixing matrices. In
Figure~\ref{fig:br} we present the branching ratios as a function of the
sgluon mass for two sets of average squark and gluino masses, and assuming
maximal mixing in the up-squark sector.  The mixed
heavy--light quark decays can dominate the sgluon decays for small
masses and decreases with larger sgluon masses.  For the maximal mixing
considered here, the decay into $t \bar{t}$ is roughly comparable to any
single one of the heavy-light decays.  Thus, combing channels together, the
branching ratio into one top (or anti-top) and a light quark are a factor of a
few times larger than that into a top pair.  For heavier squarks the
supersymmetric decay channels are typically closed in the region accessible to
the LHC. Again, the decay to one heavy and one light quark typically 
dominates.  At large sgluon masses decays into gluons dominate because the
$G$--$g$--$g$ coupling is a dimension five operator which grows with the
invariant mass of the sgluon, \ie the only scale in the process, while the
decay to quarks will be suppressed by a relative factor $m_t/\msg$.\medskip

When sgluons are pair-produced, each sgluon is as likely to decay into
a top as an anti-top.  Thus, half of the decays where both sgluons
decay into a top and a light-quark initiated jet will have same sign
tops ($tt$ or $\bar{t}\bar{t}$).  When both tops decay leptonically we
have a final state containing two light jets and either $b \ell^+ \nu
~ b \ell^{\prime+} \nu$ or a $\bar{b} \ell^- \bar{\nu} ~ \bar{b}
\ell^- \bar{\nu}$ (where $\ell$ is an electron or muon) - a striking
signature of physics beyond the Standard Model
that also arises in the context of models of top 
compositeness~\cite{Lillie:2007hd}.\medskip
 
Reference~\cite{BarShalom:2008fq} has considered 
a search for like-sign tops at the Tevatron in the
context of a model of maximal flavor violation. The authors perform a
sophisticated treatment of the backgrounds and the CDF detector
efficiencies.  While the maximal flavor violation model signal is the
result of a mixture of pair and single production of a neutral color
singlet scalar $\eta$ which interacts moderately strongly with top and
charm, the analysis only requires a pair of like sign leptons, a
$b$-tagged jet, and missing energy. Thus, our signal events are
expected to have a high efficiency with respect to the analysis
cuts, and one could get a Tevatron bound on the sgluon mass
from a similar analysis.\medskip

A single sgluon produced at the LHC can decay through its
flavor-violating interactions into a single top quark and a light jet,
similar in topology to the $s$-channel mode of single top production.
Even in the Standard model, this mode is challenging at the LHC
because of large backgrounds from $t \bar{t}$ and $t$-channel single
top~\cite{Harris:2002md}. Therefore, single sgluon production is
unlikely to be phenomenologically relevant after considering QCD
effects and backgrounds.  However, if the sgluon is heavy enough it
may be possible to use the peak in the top plus light jet invariant
mass to isolate a signal~\cite{Tait:2000sh}.
 
\begin{figure}[t]
\includegraphics[width=7cm]{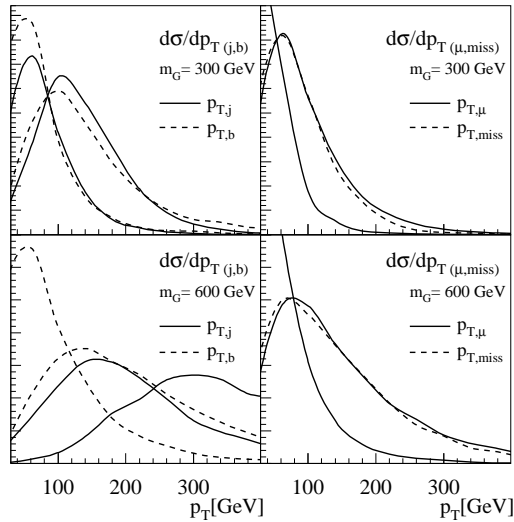} 
\caption{Transverse momentum spectra from sgluon pair production at
  the parton level on a linear scale. We show the harder/softer
  light-flavor jets, bottoms and leptons as well as the missing
  transverse momentum due to neutrinos from purely leptonic sgluon
  pairs. The sgluon masses are 300 and 600~GeV.
\label{fig:signal}}
\end{figure} 

\subsection{LHC signatures}
\label{sec:lhc}

From the discussion above we are immediately lead to study in more detail
sgluon pair production with a subsequent decay into two like-sign tops with
leptonic decays. With sizeable production rates and branching ratios above
10\% the question remains if this channel survives the LHC triggers and
acceptance cuts. In Fig.~\ref{fig:signal} we show the normalized transverse
momentum distributions of all sgluon decay particles. For both sgluon masses
of 300 and 600~GeV the light-flavor jet and bottom transverse momenta peak
above $p_T = 50$~GeV, which indicates that observing four jets and tagging two
bottoms should not be a problem. For a 300~GeV sgluon the harder jet has a
typical transverse momentum around $\ptj \sim \msg - m_t \sim 120$~GeV.
Similarly, the harder bottom can acquire $p_{T,b} \sim m_t - m_W \sim
100$~GeV. For a heavier sgluon we see that the bottom distributions hardly
change, since they are mostly determined by the top and $W$ masses. The two
light-flavor jets from the sgluon decay become significantly harder and peak
around $\ptj \sim 180$~GeV and 280~GeV, respectively. As we will see later, in
particular for heavier sgluons the decay jet can be identified unambiguously
even in the presence of QCD jets.\medskip

On the lepton side of Fig.~\ref{fig:signal} we see that the harder
lepton with a typical transverse momentum close to 100~GeV guarantees
the triggering of sgluon pair production. Moreover, such a large
transverse momentum might be useful to distinguish the sgluon pairs
from the Standard Model backgrounds, even in the same-sign lepton
case. The second lepton is comparably soft, and the missing transverse
momentum peaking around 70~GeV is unlikely to contribute to the
smoking-gun signature.\medskip

Nevertheless, from the distributions in Fig.~\ref{fig:signal} we can
see that triggering and acceptance cuts will not be a problem for the
largely background-free like-sign tops signature. Moreover, both hard
light-flavor decay jets as well as the relatively hard lepton spectrum
should help to reconstruct the sgluon mass scale, which would allow us
to gain information on the branching ratio and thereby on the flavor
structure of the MRSSM.

\begin{figure}[t]
\includegraphics[width=5.2cm]{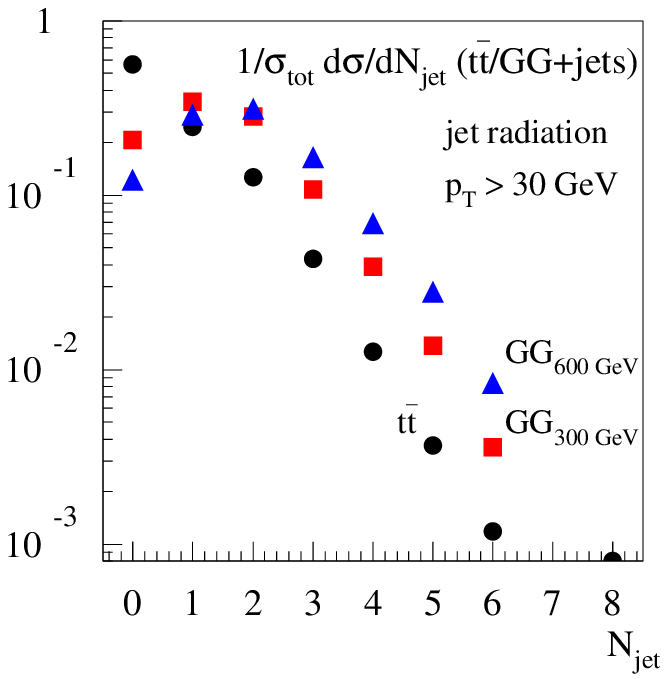} \hspace*{-0.1cm}
\includegraphics[width=5.2cm]{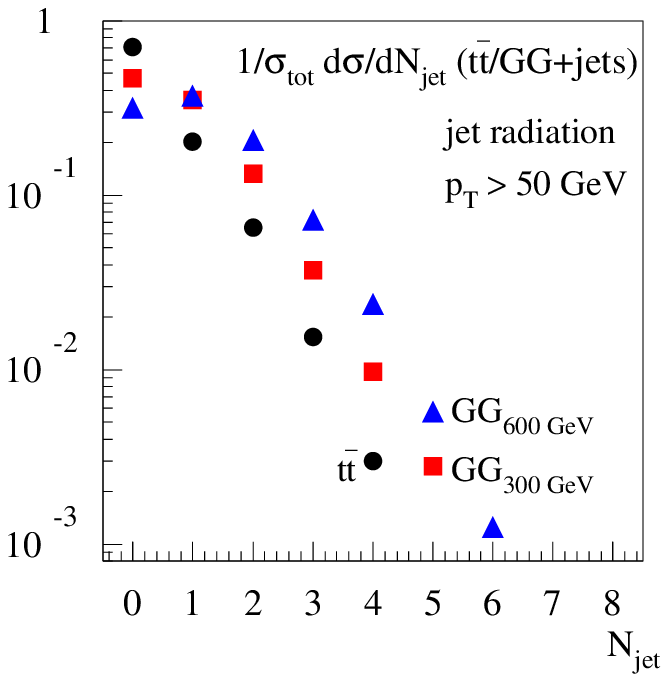} \hspace*{-0.1cm}
\includegraphics[width=5.2cm]{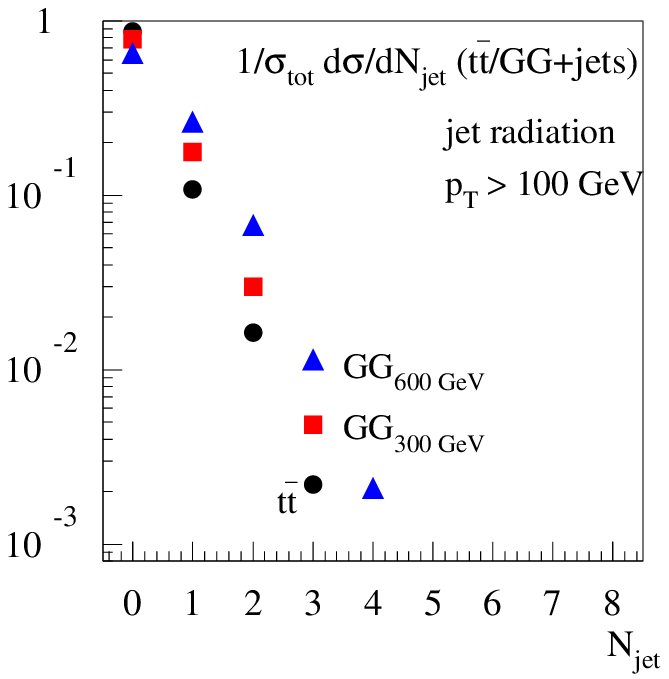} 
\caption{Number of jets produced in QCD jet radiation in addition to
  the sgluon or top decay jets. We use MLM merging as implemented in
  MadEvent and apply $p_{T,j} > 30, 50, 100$~GeV and $\Delta R_{jj} >
  0.4$ on the parton level.
\label{fig:jets}}
\end{figure} 

\subsection{Sgluons and QCD jets}
\label{sec:qcd}

Because the light-flavor decay jets from the sgluons are crucial for the
analysis described above, we have to answer the question if this jet can be
identified in the jet-rich LHC environment. This question becomes even more
relevant once we try to search for the opposite-sign lepton channel or attempt
to determine the sgluon mass from hadronic top decays. Recently, there have
been theoretical developments which allow us to simulate QCD jet radiation
over the entire transverse momentum range of these additional jets and
including high jet multiplicities.\medskip

We simulate additional QCD jets in the signal process $pp \to GG+X$ using the
MadEvent~\cite{mlm_me} implementation of the MLM scheme. In particular because
the sgluons in the final state provide a hard scale for the process we do not
expect the MLM results to differ from a corresponding CKKW
analysis~\cite{ckkw,compare_them,my_tasi}. In particular for heavier sgluon
masses the QCD activity should be dominated by collinear parton-shower
effects~\cite{skands}, but the MLM scheme now allows us to consistently treat
top backgrounds and the sgluon signal for different masses. For this
simulation we avoid introducing a supersymmetric shower including sgluon
splittings, which would be required to include final-state radiation. Because
of the lack of a collinear enhancement we know that final-state radiation will
not contribute strongly to QCD jet radiation. Therefore, we only include
initial-state radiation which is universal for different heavy new-physics
states produced at the LHC.\medskip

In Fig.~\ref{fig:jets} we show the number of QCD jets in sgluon pair events at
the LHC. Apart from a crucial minimal jet separation of $R_{jj} > 0.4$ we
apply only a varying transverse momentum cut of 30, 50 and 100~GeV on the
radiated jets. In the left panel we see that while top quarks most likely come
with no additional jet from initial-state radiation, a 300~GeV sgluon will
most likely be accompanied by one and a 600~GeV sgluon by two additional QCD
jets. This is an effect of the hard factorization scale in the process which
determines the size and the maximum range of the collinear enhancement of
initial-state radiation. When we increase the minimum $\ptj$ to 50~GeV the
typical number of additional jets drops by roughly one, but in particularly
heavy states still come with zero, one, or two jets at roughly the same rate.
Only the three additional jets channel is suppressed to the 10\% level, where
this quantitative result should be taken with a grain of salt without a tuned
parton shower for the LHC. Finally, a cut of at least $\ptj > 100$~GeV gets
rid of additional jets in roughly two thirds of the events and allows us to
use light-flavor decay jets in the analysis.\medskip
 
\begin{figure}[t]
\includegraphics[width=7cm]{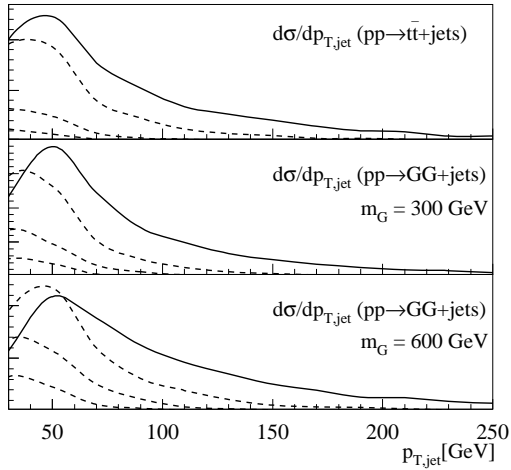} 
\caption{Transverse momentum of the leading and the three sub-leading
  QCD jets radiated in sgluon and top pair production. The different
  curves are normalized according to the relative rates in
  Fig.~\ref{fig:jets}. The leading jets is shown as a solid line.
\label{fig:merging}}
\end{figure} 

The normalized transverse momentum distributions for the radiated jets are shown
in Fig.~\ref{fig:merging}. The different areas under the curves for the four
leading jets reflect the fact that only a fraction of events actually show
such jets. Again, we see that from the top pairs to the 600~GeV sgluons the
situation changes qualitatively: in the bottom panel the
curves for the leading and the sub-leading are similar, and as long as we stay
below $\ptj$ even a third QCD jet is very likely to appear. In a way, the
crossing point between the two hardest jet indicates a $\ptj$ range below
which we should not rely on QCD jets being rare or suppressed. In other words,
50~GeV jets should not be used as part of the signal unless we have a way to
identify decay jets, while jets with $\ptj > 100$~GeV are comparably safe. Of
course, this insight is not new --- for example, careful squark and gluino
analyses for the LHC have always been designed that
way~\cite{sq_go,osland}.\medskip

The good news of this QCD analysis is that jet activity will actually be
helpful to distinguish sgluon pairs from $t\bar{t}$ backgrounds. First, top
pair events are quite unlikely to come with a QCD jet of $\ptj > 100$~GeV,
which we have seen in Sec.~\ref{sec:lhc} is typical for the signal's decay
jets. Secondly, as long as we require $\ptj > 100$~GeV for the decay jets,
even a 600~GeV sgluon will hardly come with such hard QCD jets. In general, if
we consider something like the total visible mass as an observable to
distinguish the sgluon-pair signal from top pairs, Fig.~\ref{fig:jets} shows
that QCD radiation will improve this handle. Based on these results, we can
estimate that for sgluon branching ratios into $t\bar{q}$ of order 10\%, we
can expect an LHC discovery reach of up to roughly $\msg \sim 1$~TeV.\medskip

The bad news is that for the reconstruction of the sgluon mass we
could utilize the semileptonic sample. From Fig.~\ref{fig:signal} we
can guess that the jet from the hadronic top decay should be
comparable to the lepton spectrum, which does not guarantee that both
of them even lie above $\ptj > 50$~GeV, where the safe region would
really only start at $\ptj > 100$~GeV, as seen in
Fig.~\ref{fig:jets}. Reconstructing the sgluon including a hadronic
top decay is unlikely to succeed in a realistic QCD environment. As mentioned
above, single sgluon production might rely on such a reconstruction for a
side-bin analysis against the theoretically notorious single-top background,
so QCD effects are unlikely to help with this problem.

\section{Conclusions}
\label{sec:conclusions}

The MRSSM, with a continuous $R$ symmetry,
is an interesting alternative to the MSSM, which provides a novel
solution to the tension between supersymmetry at the TeV scale
and measurements from the flavor sector.  A key feature of any
$R$--symmetric model is the promotion of gauginos to Dirac fermions,
which automatically implies the existence of an additional color-octet 
scalar with Standard-Model type $R$ charge, the sgluon.  Sgluons
necessarily have tree-level couplings to gluons, gluinos, and squarks,
which further induce couplings to quarks and (single sgluon coupling to) 
gluons through loops.\medskip

Sgluons have a large color charge and can be copiously produced at the 
LHC.  Their decays can include squarks, gluinos, gluons, and quarks.
This last decay, through the large squark mixing which is the hallmark of
the MRSSM, can violate flavor, leading to large branching ratios into
a top and a light quark.
Pair production with a subsequent decay to
light-sign top quarks appears to be the most promising search channel
at the LHC. We have computed the relevant LHC cross sections and
branching ratios, with special focus on jet activity in this
essentially background-free signature.\medskip

Properly simulated QCD effects turn out to be helpful with regard to
the sgluon pair analysis: additional jet radiation in the signal will
typically create more hard jets, adding to the two hard decay jets
from the sgluon pair. While the QCD jets are too soft to be mistaken
for decay jets they create a generally harder signal event, while jet
radiation for the already softer top-pair background makes this
general feature even more prominent. In the light of this result, a
search for sgluon pairs with a $t\bar{t}$ final state might be
feasible.

For single sgluon production the QCD-induced background uncertainty is
potentially dangerous. The obvious way to tell apart signal and
background would be to look for a peak in the top--jet invariant mass,
with a hadronically decaying top quark. However, QCD jet radiation can
be expected to lead to a sizeable combinatorical background to the $W$
decay jets, which is worsened by the generically higher scale of the
signal process.

\section*{Acknowledgments}

The authors are grateful for conversations with Beate Heinemann, Ben
Kilminster, Charles Plager, Dave Rainwater, and Daniel Whiteson. For
the QCD simulations we are hugely grateful to Johan Alwall for
teaching us Madevent and jet merging. We also would like to thank the
Aspen Center for Physics for providing such a stimulating environment
which inspired this study.  
We thank Claude Duhr and Fabio Maltoni for helping us implement the correct
color structure for the $G$-$G$-$g$-$g$ vertex in MadEvent, 
which is not properly gauge invariant by default.
We are grateful to the authors of \cite{Choi:2008ub} for detailed comparisons
with part of our results and
correcting a sign error in our original Feynman rules.  Our phenomenological
results have not changed significantly.
Research at Argonne National Laboratory is
supported in part by the Department of Energy under contract
DE-AC02-06CH11357.

\appendix

\section{Loop Induced Couplings of Sgluons}

In this appendix, we present the expressions for sgluon coupling to 
quarks and gluons induced
at one loop.  The results are presented in terms of the scalar integral 
functions of
Passarino and Veltman~\cite{Passarino:1978jh}, normalized such that 
the measure is $d^n q /(i\pi^2)$.

The sgluon coupling to quark $q^i$ and anti-quark $\bar{q}^j$ receives 
contributions from both right- and left-handed squarks.  In the MRSSM, 
right- and left-handed squarks do not mix with each other, so even with 
large squark mixing, they form two distinct sectors.  We are interested in
two cases.  The first has $m_i \gg m_j$, for which we can approximate,
\begin{alignat}{5}
&f_{q,i}^{R} = \sqrt{2} \; 
              \frac{\msg^2}{\msg^2-m_i^2} \times \notag \\
&
 \Big\{ \;  N_c \left[
    B \left(m_i^2; \mgo, \msq \right) 
  - B \left(\msg^2; \mgo, \mgo \right) 
  + \left( \mgo^2 - \msq^2  \right)
    C \left( m_i^2, 0, \msg^2; \mgo, \msq, \mgo \right) \right]
 \notag \\
& 
 - \frac{1}{N_c} \left[
    B \left(m_i^2; \mgo, \msq \right) 
  - B \left(\msg^2; \msq, \msq \right)
  + \left( \mgo^2 - \msq^2 \right)
    C \left( m_i^2, 0, \msg^2; \msq, \mgo, \msq \right) \right]
\Big\}
\end{alignat}
where the left-handed squarks contribute.  And,
\begin{alignat}{5}
&f_{q,i}^{L} = -\sqrt{2} \; 
              \frac{\msg^2}{\msg^2-m_i^2} \times \notag \\
&
 \Big\{ \;  N_c \left[
    B \left(m_i^2; \mgo, \msq \right) 
  - B \left(\msg^2; \mgo, \mgo \right) 
  + \left( \mgo^2 - \msq^2 + m_i^2 - \msg^2 \right)
    C \left( m_i^2, 0, \msg^2; \mgo, \msq, \mgo \right) \right]
 \notag \\
& 
 - \frac{1}{N_c} \left[
    B \left(m_i^2; \mgo, \msq \right) 
  - B \left(\msg^2; \msq, \msq \right)
  + \left( \mgo^2 - \msq^2 \right)
    C \left( m_i^2, 0, \msg^2; \msq, \mgo, \msq \right) \right]
\Big\}
\end{alignat}
 where the right-handed squarks are running in the loops.  The chiral
 couplings are a consequence of eq.(\ref{eq:susy_qcd}) and will be
 different for different admixtures of $G$ and $G^*$. 
The $f_{q,j}^{(L/R)}$ multiply $m_j$ and thus can be neglected.
The second case has $m_i = m_j$.  The relevant quantities are
$f_{q,i}^{(R/L)} - f_{q,j}^{(R/L)}$ given by,
\begin{alignat}{5}
&f_{q,i}^{(R/L)}  - f_{q,j}^{(R/L)} = \sqrt{2} \; 
                            \frac{\msg^2}{\msg^2-4m_i^2} \times \notag \\
& 
 \left\{ 2 N_c \left[
    B \left(m_i^2; \mgo, \msq \right) 
  - B \left(\msg^2; \mgo, \mgo \right) 
  + \left(\mgo^2 - \msq^2 + m_i^2 - \frac{\msg^2}{2} \right)
    C \left( m_i^2, m_i^2, \msg^2; \mgo, \msq, \mgo \right) \right]
 \right. \notag \\
& \left. 
 - \frac{2}{N_c} \; \left[
    B \left(m_i^2; \mgo, \msq \right) 
  - B \left(\msg^2; \msq, \msq \right)
  - \left( \mgo^2 - \msq^2 + m_i^2 \right)
    C \left( m_i^2, m_i^2, \msg^2; \msq, \mgo, \msq \right) \right]
\right\}
\end{alignat}
where the right-handed squarks contribute to $f_L$ and vice-versa.
The sgluon coupling to gluons for one squark flavor is
\begin{equation}
\lambda_g = 2 \sqrt{2} \; \sum_{\tilde{q}} \;
  \left[  m_{\sq_L}^2 C \left( 0, 0, \msg^2; m_{\sq_L}, m_{\sq_L}, m_{\sq_L} \right)
- m_{\sq_R}^2 C \left( 0, 0, \msg^2; m_{\sq_R}, m_{\sq_R}, m_{\sq_R} \right)
\right]
\label{eq:formfac}
\end{equation}
summed over all of the $n_f$ squark flavors.  From the
naive computation, the color factor of this coupling includes the
antisymmetric $f_{abc}$ as well as the symmetric $d_{abc}$. However, due to
the symmetry structure of eq.(\ref{eq:formfac}) only $d_{abc}$ survives. For
the same reason, the gluino loop contribution with its only color structure
$f_{abc}$ cancels completely.

\end{document}